# Local characterization of ferromagnetic properties in ferromagnet/superconductor bilayer by Point Contact Andreev Reflection Spectroscopy

*Filippo Giubileo\*,[a,b,‡], Francesco Romeo[b,‡], Roberta Citro[b,a], Antonio Di Bartolomeo[b], Carmine Attanasio[b,a], Carla Cirillo[a,b], Albino Polcari[c], Paola Romano[c,a]*

[a] CNR-SPIN Salerno, via Giovanni Paolo II 132, 84084 Fisciano (SA), Italy

[b] Dipartimento di Fisica "E.R. Caianiello", Università degli Studi di Salerno, via Giovanni Paolo II 132, 84084 Fisciano (SA), Italy

[c] Dipartimento di Scienze e Tecnologie, Università degli Studi del Sannio, via Port'Arsa 11, Benevento, Italy

ABSTRACT

We realized point contact spectroscopy experiment on ferromagnet/superconductor bilayers. Differential conductance curves show several features that we explained within Bogoliubov-de Gennes formalism considering the presence of two interfaces in the normal-metal-tip/ferromagnet/superconductor device. We demonstrate that such configuration is suitable as local probe of the spin polarization and thickness of ferromagnetic layer, directly on bilayer areas. This is due to the high sensitivity of the Andreev surface states to the physical properties of the ferromagnetic interlayer.

Spin polarization (P) represents an intrinsic parameter that characterizes a ferromagnet measuring the spin imbalance for the occupied electronic states. Experimentally, P can be determined by photoemission spectroscopy[1] (PS) as well as by spin-dependent spectroscopy on magnetic tunnel junctions[2] (MTJ), but both methods have important drawbacks: PS has limited energy resolution (few meV) and spatial sensitivity (few Angstrom of the surface) while MTJs need high quality fabrication process to get planar structures with uniform thin insulating barrier and a setup to apply high magnetic fields. Less than twenty years ago, De Jong and Beenakker[3] proposed the possibility to measure P by means of Point Contact Andreev Reflection (PCAR) spectroscopy exploiting the Andreev reflection (AR) process at the metal/superconductor interface for which an incoming electron with energy less than the superconducting energy gap is retroreflected in the metal as a hole with opposite spin, while a Cooper pair enters the superconductor[4]. If the metal is a ferromagnet, the probability for AR is reduced and (transport) polarization can be obtained from the study of the differential conductance spectra G(V) (conductance vs voltage). The main advantage is clearly related to the simplicity to realize a superconductor/ferromagnet micro-constriction by pushing a tip on a sample, avoiding fabrication process of tunnel junctions and without application of external magnetic field. On this idea, first experimental evidences of measuring P in ferromagnet materials by PCAR have been independently reported in 1998 by R.J. Soulen et al.[5] and by S.K. Upadhyay et al.[6]. Later, this technique has been used to characterize a large number of ferromagnetic metals[5-8] (Fe, Co, Ni), alloys[5,9] (permalloy $Ni_xFe_{1-x}$), manganites[5,10-11] ($La_{1-x}Sr_xMnO_3$), ruthenates[12-13] ($SrRuO_3$) and half metals[5,14-15] ($CrO_2$). Nowdays, PCAR experiment is well established as a valid method to measure P.

From a theoretical point of view, a simple approach by Strijkers et al.[7] gives a generalization of the BTK model[16] to spin polarized materials by considering the current flowing in a ferromagnet/superconductor (F/S) contact as $I = (1 - P) \cdot I_u + P \cdot I_p$, with $P$ the spin polarization in F and $I_p$ and $I_u$ the fully polarized and fully not polarized current, respectively (the AR process being zero for the polarized case). Moreover, by considering the presence of a weak superconducting layer at the interface due to proximity effect, this model succeeded in some cases to fit conductance dips often experimentally observed at energies close to the gap energy. Another model to describe AR at the F/S interface by F. Peréz-Willard et al.[17] takes into account two spin-dependent transmission coefficients for the majority and minority carriers in the ferromagnet. Both models have been widely applied in order to extract spin polarization in several experiments in which point contact is realized between a ferromagnetic material and a superconductor.

In this Letter we extend the use of point contact technique to characterize F/S bilayers and extract direct local information about spin polarization and thickness of the ferromagnetic layer. We have developed a theoretical model within a Bogoliubov-de Gennes (BdG) formalism[18] taking into account the presence of two interfaces, tip/sample (N/F) and F/S (in the bilayer). We then applied such model to analyse experimental results obtained in PCAR experiment by pushing a gold tip on the ferromagnetic side of a PdNi/Nb bilayer. We demonstrate that in this configuration, PCAR can give extremely precise estimation of transport spin polarization as well as of the local ferromagnet thickness, the high sensitivity being due to the strong dependence of the surface (Andreev) bound states on such physical properties.

*Model*. We adopt a Bogoliubov-de Gennes formalism to describe the PCAR setup. Accordingly, the wave function $\phi(r)$, describing an excitation of energy $E$ in the tip, in the ferromagnetic layer or in the superconducting substrate, is derived by solving the eigenvalues problem given by ($x \neq 0, d$)

$$\begin{bmatrix} H(r) & \Delta(r) \\ \Delta^+(r) & -H^*(r) \end{bmatrix} \phi(r) = E\phi(r). \tag{1}$$

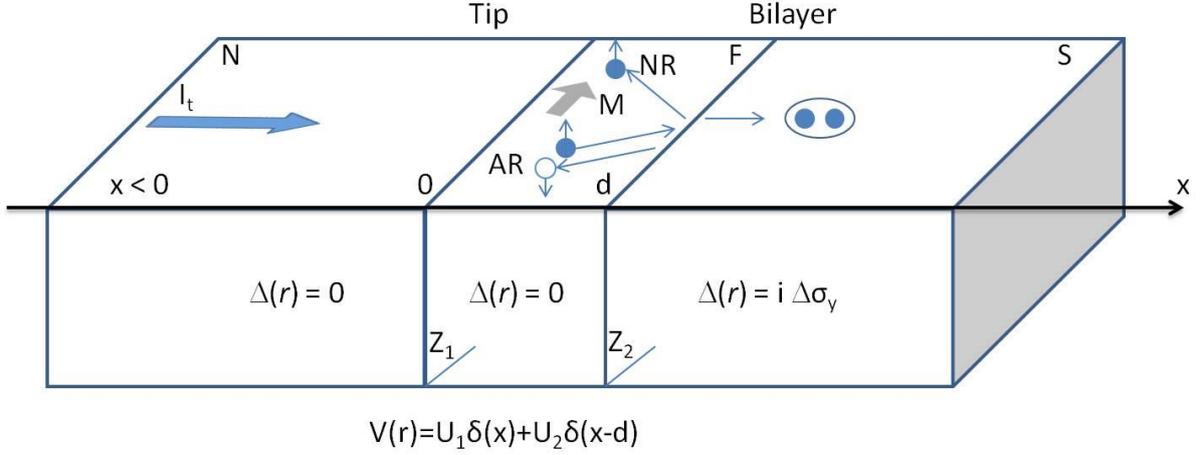

**Figure 1.** Schematic representation of theoretical model: transport current $I_t$ flows from metallic tip through N/F (parameterized by $Z_1$) and then F/S ($Z_2$) interfaces, barrier strengths depending on the scattering potential $V(r)$. Ferromagnetic region ($0 < x < d$) has magnetization $M$ perpendicular to transport direction. Superconducting pairing potential, existing in S, causes AR and normal reflection only at F/S interface.

The tip region, characterized by $x < 0$, does not present superconducting correlations and thus we set $\triangle(r) = 0$ (see Figure 1), while the tip quasi-particle Hamiltonian is assumed to be of free-particle form $H(r) = -\frac{\hbar^2 \nabla^2}{2m} - E_F \equiv H_0(r)$. The thin ferromagnetic layer ($0 < x < d$) is modeled by setting $\triangle(r) = 0$ and by adding to the free-particle Hamiltonian $H_0(r)$ a Zeeman energy term, $-g\mu_B M \sigma_z \equiv -E_F h \sigma_z$, describing a magnetization $M$ belonging to the $y - z$ plane (i.e. the magnetic easy plane) orthogonal to the transport direction, i.e. the $x$-direction. The superconducting region ($x > d$) is described by a homogeneous pairing potential $\triangle(r) = i\triangle\sigma_y$, while the quasi-particle Hamiltonian $H_0(r)$ is assumed. The Fermi velocities mismatch among the different regions and the non-ideality of the interfaces are modeled by using a scattering potential $V(r) = U_1\delta(x) + U_2\delta(x - d)$ to be added to the single particle Hamiltonian $H_0(r)$, $\delta(x)$ being the Dirac delta function. We assume the translational invariance of the problem along any direction belonging to the $y - z$ plane which implies the conservation of the linear momentum $\hbar \mathbf{k}_\parallel \equiv \hbar(0, k_y, k_z)$ parallel to the interface. The wave function in each region can be written in the form $\phi(r) = e^{i(k_y y + k_z z)}\psi(x|E, \mathbf{k}_\parallel)$ leading to an effective one-dimensional problem for $\psi(x|E, \mathbf{k}_\parallel)$, being the energy $E$ and $\mathbf{k}_\parallel$ conserved quantum numbers during a scattering event. Once the wave functions $\psi_t(x|E, \mathbf{k}_\parallel)$, $\psi_f(x|E, \mathbf{k}_\parallel)$ and $\psi_s(x|E, \mathbf{k}_\parallel)$ describing, respectively, the tip, the magnetic layer and the superconducting substrate have been expressed in terms of eigenfunctions associated to the eigenvalues problem given in Equation 1, the scattering coefficients are determined by imposing the boundary conditions: (i) $\psi_t(x = 0|E, \mathbf{k}_\parallel) = \psi_f(x = 0|E, \mathbf{k}_\parallel)$, (ii) $\psi_s(x = d|E, \mathbf{k}_\parallel) = \psi_f(x = d|E, \mathbf{k}_\parallel)$, (iii) $\partial_x \psi_f(x|E, \mathbf{k}_\parallel)\big|_{x=0} - \partial_x \psi_t(x|E, \mathbf{k}_\parallel)\big|_{x=0} = k_F Z_1 \psi_t(x = 0|E, \mathbf{k}_\parallel)$, (iv) $\partial_x \psi_s(x|E, \mathbf{k}_\parallel)\big|_{x=d} - \partial_x \psi_f(x|E, \mathbf{k}_\parallel)\big|_{x=d} = k_F Z_2 \psi_f(x = d|E, \mathbf{k}_\parallel)$, where $k_F$ indicates the Fermi wave vector, while $Z_{1/2} = 2mU_{1/2}/(\hbar^2 k_F)$ represents the BTK parameter describing the interface properties. The current $I_t$ flowing through the constriction can be expressed via the AR coefficients $a_{\sigma'\sigma}(E, \mathbf{k}_\parallel)$ and the normal reflection coefficients $b_{\sigma'\sigma}(E, \mathbf{k}_\parallel)$ defining the tip wave function $\psi_t = \psi_{e\sigma}^{in} + \sum_{\sigma'} b_{\sigma'\sigma} \psi_{e\sigma'}^{out} + \sum_{\sigma'} a_{\sigma'\sigma} \psi_{h\sigma'}^{out}$. Here $\psi_t$ is decomposed into incoming (in) or outgoing (out) electron-like ($\psi_{e\sigma}^{in/out}$) and hole-like ($\psi_{h\sigma}^{in/out}$) modes having spin projection $\sigma\hbar/2$, with $\sigma = \pm 1$. The experimentally measured differential conductance has to be compared with $G(V) = \frac{dI_t}{dV}$ where:

$$I_t(V) \propto \frac{A}{(2\pi)^2} \int dE\, d^2\mathbf{k}_\parallel [2 + \sum_{\sigma\sigma'}|a_{\sigma'\sigma}(E, \mathbf{k}_\parallel)|^2 - \sum_{\sigma\sigma'}|b_{\sigma'\sigma}(E, \mathbf{k}_\parallel)|^2](f(E - eV) - f(E)),$$

$f(E)$ is the Fermi-Dirac distribution and $A$ represents the junction cross section. The above expression can be rewritten in terms of angular integration over the incidence angles $(\theta, \varphi)$ by changing the double integral variables taking the modulus $k(E) = \sqrt{2mE}/\hbar$ of the wave vector as fixed, i.e. $\int d^2 \mathbf{k}_\parallel \to \int k^2(E) (\sin \varphi)^2 \cos \theta \, d\theta d\varphi$. In the PCAR experiment the voltage bias $eV$ ranges from zero to few times the superconducting gap $\Delta$, as a consequence, in the relevant energy window $[\mu, \mu + eV]$ around the chemical potential $\mu$, the wave vector $k(E)$ is well approximated by the constant value $k_F$, being the corrections to the leading term of order of $\frac{eV}{\mu} \sim \frac{\Delta}{\mu} \approx 10^{-3}$. Thus the central quantity of our analysis can be written as:

$$I_t(V) \propto \frac{k_F^2 A}{(2\pi)^2} \int dE \, d\Omega \, [2 + \sum_{\sigma\sigma'} |a_{\sigma'\sigma}(E, \theta, \varphi)|^2 - \sum_{\sigma\sigma'} |b_{\sigma'\sigma}(E, \theta, \varphi)|^2](f(E - eV) - f(E)),$$

where we introduced the notation $d\Omega \equiv (\sin \varphi)^2 \cos \theta \, d\theta d\varphi$, while the angular integration is performed over $\theta \in [-\pi/2, \pi/2]$ and $\varphi \in [0, \pi]$. We notice that the factor $\frac{k_F^2 A}{(2\pi)^2}$ is related to the number of transverse modes which participate in the charge transport. Once the differential conductance $G(V)$ is determined using the above relation, it is normalized with respect to the differential conductance of the junction at high bias, i.e. using the value $G_{NN} = G(V)|_{eV \gg \Delta}$. The quantity $G(V)/G_{NN}$ is directly compared with the experimental data.

*Experiment*. The bilayers measured in this study were grown in-situ by a three-target ultra high vacuum dc magnetron sputtering on $Al_2O_3$ substrates in Argon pressure (few μbar) depositing first a 40 nm thick Nb layer and then a 4 nm thick $Pd_{0.84}Ni_{0.16}$ layer. The critical temperature of the bilayer $T_c^{bilayer}$ was checked by resistive transition measurement and it has been compared with the same parameter of a twin Nb film ($T_c^{Nb} = 8.2 \, K$, without the ferromagnetic layer on top) resulting about 1K lower.

PCAR experiments have been performed by pushing a mechanically etched gold tip on the ferromagnetic side of the PdNi/Nb bilayer. The tip is installed on a screw driven chariot in order to allow gentle approach to the sample surface. The measuring inset is directly introduced in a liquid helium cryostat, with the device exposed to helium atmosphere. Conventional four-probe technique (see inset of Figure 1) has been applied in order to measure current-voltage (I-V) characteristics in the temperature range between 4.2K and 10K (i.e., above the Nb critical temperature). Differential conductance spectra (G-V) are obtained by numerical derivative of the I-V curves. Several different contact resistances have been obtained in the range 2Ω-10Ω by simply varying the position and the pressure of the tip on the sample.

The transport regime for such contacts can be easily estimated to be ballistic or diffusive by using Wexler's formula[19]

$$R_{PC} = \frac{4\rho l}{3\pi a^2} + \frac{\rho}{2a}$$

in which the first term gives the Sharvin resistance[20] describing the ballistic regime and the second one is the Maxwell[21] resistance describing the diffusive regime. The dominating term will depend on the contact dimension $a$, the resistivity $\rho$ of the sample and the mean free path $l$ of the charge carriers. For $\rho = 13 \, \mu\Omega \, cm$ (as resulting by direct measurements) and considering that $\rho l = 3.72 \times 10^{-6} \mu\Omega \, cm^2$ for niobium[22-23], it comes out that the minimum contact dimension in our junctions is $a \approx 8$ nm to be compared with the mean free path $l \approx 3$ nm. The ratio $l/a < 1$ gives indication for diffusive transport. However, the extention of BTK theory to diffusive regime has been proven[24] to be successful in correctly identify the effect of spin polarization on the conductance spectra with respect the effects due to the diffusive transport. Moreover, it has been demonstrated[25] that the application of ballistic model to analyze PCAR spectra obtained in the diffusive regime will allow an estimation of the spin polarization with an error below 3%. At the same time, the barrier parameter Z evaluated in a ballistic model will be systematically larger than what obtained in a diffusive model (with a variation of about 0.5-0.6) due to the fact that the parameter Z should include either the ballistic barrier strength (as expected in the BTK theory) and other physical effects (diffusion, velocity and/or mass mismatch). In Figure 2, we show a variety of normalized conductance spectra (conductance is expressed as $G(V)/G_{NN}$ while energy scale, eV/$\Delta_{Nb}$, is normalized to the Nb energy gap) obtained at T=4.2 K. At a first qualitative analysis, these data can appear quite puzzling: Zero Bias Conductance Peak (ZBCP) higher than 2 (i.e., $G(V=0)/G_{NN} > 2$, where $G(V=0)$ is the conductance at zero bias and $G_{NN}$ is the conductance at high bias) appears in many spectra; two conductance dips, one fixed at the niobium gap energy, another at lower (not fixed) energy, are always present; conductance maxima within gap energy appear with different intensity in the various spectra. To quantitatively analyze the conductance curve reported in Figure 2a-2d, experimental data (empty circles) are compared to theoretically calculated spectra (solid lines) according to the model introduced in the previous section: the result is satisfactory with all features properly reproduced. We used as fitting parameters the barrier strength $Z_1$ (describing the N/F interface), the barrier strength $Z_2$ (describing the F/S interface), the thickness parameter $r$ (describing the thickness of the ferromagnetic layer according to the formula $r = k_F^{Nb} \cdot d$, where $k_F^{Nb} \approx 11.8\ nm^{-1}$ is the Fermi momentum[26] and $d$ is the real thickness), and the spin polarization $h$; we do not consider as fitting parameter the niobium superconducting energy gap and the effective temperature that we fixed at the values $\Delta_{Nb}$= 1.5 meV and $T_{eff} = 0.7$ K. We notice that the effective temperature has been fixed at a value sensibly lower than the bath temperature ($T_{bath}$=4.2K): this discrepancy will be discussed in the next section.

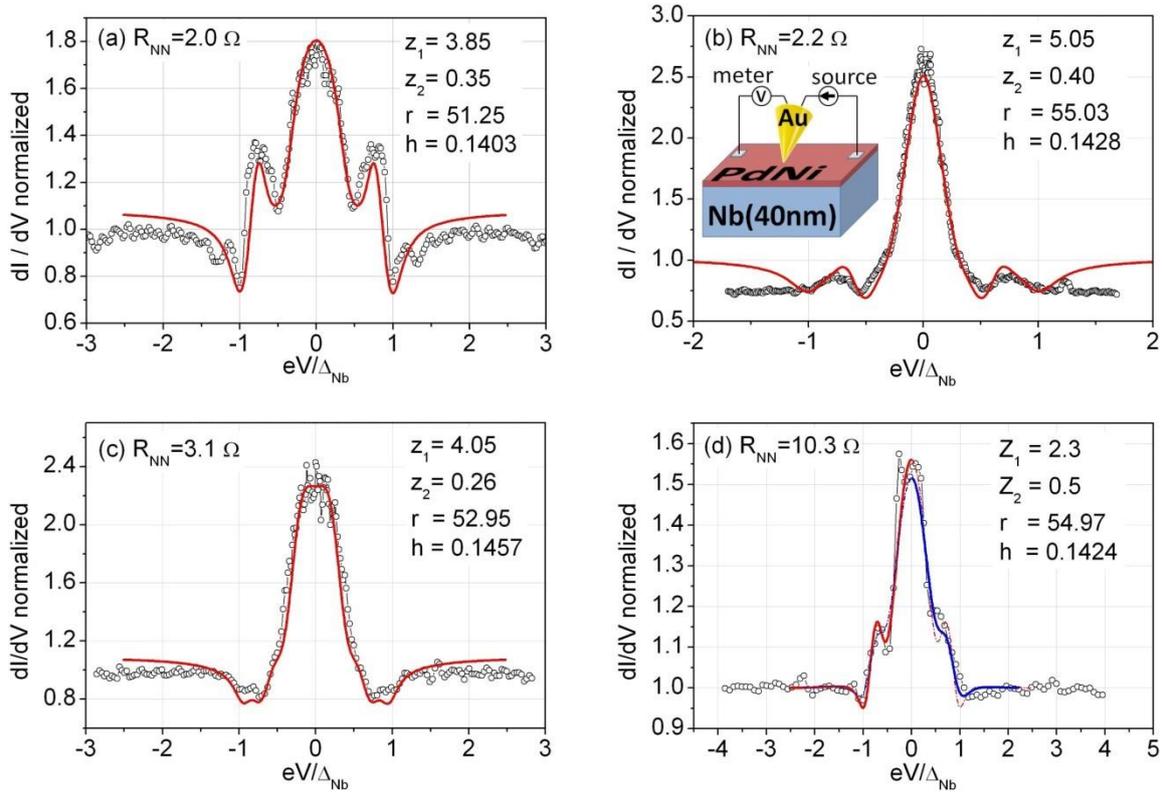

**Figure 2.** Differential conductance spectra measured at low temperature (T = 4.2 K) on contacts realized by pushing an Au tip on a PdNi/Nb bilayer. Inset in (b) shows a scheme of the setup. Different contacts are classified by $R_{NN}$, i.e. the high bias resistance of the device. Experimental data (empty circles) are normalized to $G_{NN}=1$ (where $G_{NN} = 1/R_{NN}$ is the high bias conductance) and compared to curves (solid lines) resulting from the theoretical model discussed above. In each plot are listed the parameters used in the model to reproduce the data: $Z_1$ and $Z_2$ are the barrier height of the tip(Au)/ferromagnet(PdNi) interface and of the ferromagnet(PdNi)/superconductor(Nb) interface, respectively; $r$ is related to the ferromagnet thickness $d$ via the equation $r = k_F^{Nb} \cdot d$; $h$ is the polarization of the ferromagnetic layer. All fits are performed by considering a temperature value $T_{eff} = 0.7$ K well below the bath temperature of 4.2 K. (d) The asymmetry of the spectrum is reproduced by simply assuming a higher temperature for the positive energy side $T_{eff} = 1.1$ K.

Fitting parameters used to reproduce experimental data are reported in Figure 2 for each plot. All conductance spectra are characterized by a large $Z_1$ value ($2.3 < Z_1 < 5.1$) indicating a low transparency of the contact between the gold tip ant the PdNi layer; at the same time, significantly lower $Z_2$ values are always found ($0.26 < Z_2 < 0.50$) as expected for in-situ fabricated interface Nb/PdNi. More interestingly, the other two parameters involved in the fitting procedure take on values in narrow intervals, $51.2 < r < 55.1$ and $0.140 < h < 0.146$ from which it is possible to give an extimation of the ferromagnet thickness $4.3\text{nm} < d < 4.6\text{nm}$ in the various sample positions and of the corresponding spin polarization $14.0\% < P < 14.6\%$. We will discuss in the following section on the precision of such estimation.

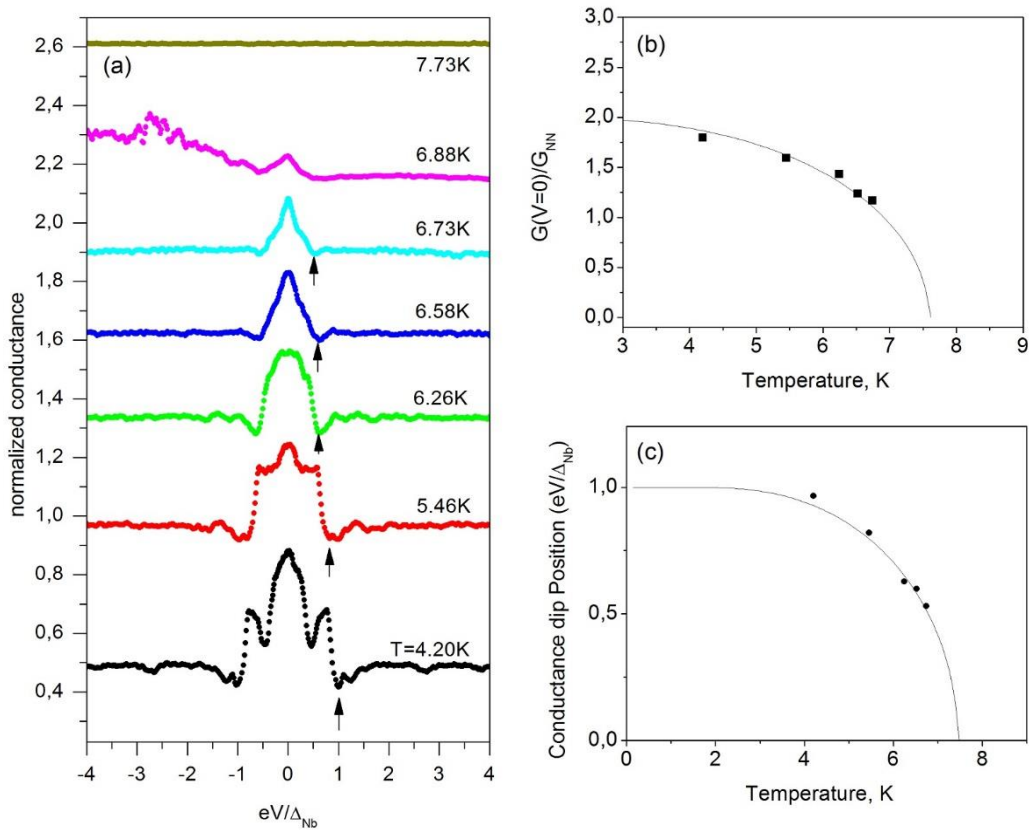

**Figure 3.** Temperature dependence of the conductance spectra measured for the contact of Figure 2a. (a) Spectra have been shifted for clarity. Black arrows identify the position of the conductance dip at the gap edge. (b) Temperature evolution of the relative amplitude of the zero bias conductance $G(V = 0)/G_{NN}$ is compared with the theoretical BCS behavior of $\Delta(T)$. (c) Temperature dependence of the Nb energy gap as extracted by the experimental data.

In order to verify that conductance features are strictly related to the superconductivity of Nb, we performed complete temperature dependence of the conductance spectra. We show in Figure 3 that for T = 7.7 K the device is not anymore superconducting and all conductance features are washed out. Moreover, the conductance dip position (black arrows in Figure 3a) and the amplitude of the zero bias peak both correctly follow the expected BCS behaviour for $\Delta(T)$.

*Discussion.* The N-F/S configuration of the point contact experiment allowed to measure several different conductance spectra, as reported in Figure 2. The use of a theoretical model based on BdG formalism gives a complete explanation of all conductance features measured at low temperatures. In Figure 4 we show the effect of the variation of the model parameters on the conductance curves. In all plots of Figure 4 the solid (black) spectrum corresponds to the calculated curve of Figure 2a whose parameters are $Z_1=3.85$, $Z_2=0.35$, $r=51.25$, $h=0.1403$. The four plots of Figure 4 are then obtained by keeping fixed three parameters and allowing only one parameter to vary in a range. Thus, Figure 4a for $0<Z_1<8$, Figure 4b for $0<Z_2<8$, Figure 4c for $48<r<58$, Figure 4d for $0.135<h<0.145$ are obtained.

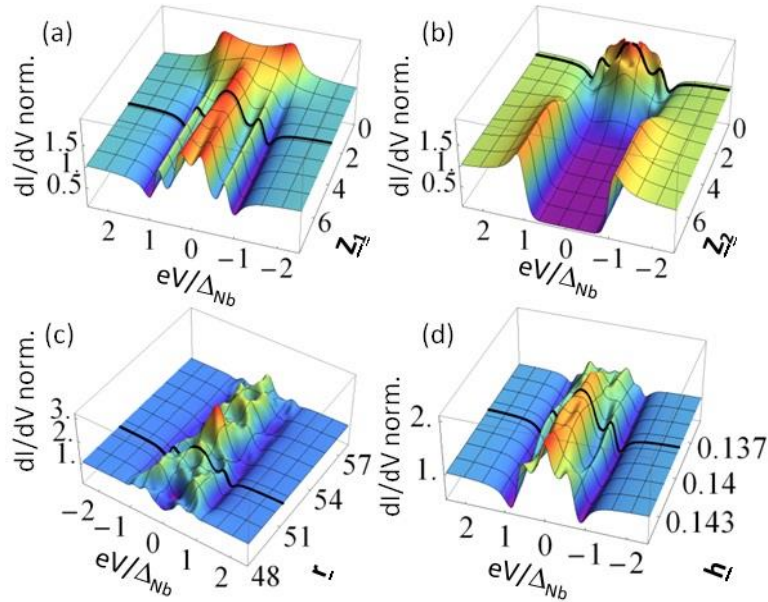

**Figure 4.** Evolution of the differential conductance spectra calculated within our theoretical model. The black spectrum evidenced in each plot corresponds to Figure 2a ($Z_1=3.85$, $Z_2=0.35$, $r=51.25$, $h=0.1403$). Different plots are obtained by varying only one parameter (a) $Z_1$, (b) $Z_2$, (c) $r$, (d) $h$ and keeping the other three parameters fixed at the value of Figure 2a.

A completely transparent contact ($Z_1 = 0$) results in a simple spectrum with two maxima close to the gap energy. For increasing $Z_1$, there is a slow evolution of the spectra and two conductance minima appear exactly at $\Delta_{Nb}$ and a ZBCP arises as well as other two maxima within energy gap. For largest $Z_1$ values, the zero bias peak further splits. Differently, by varying $Z_2$ there is a fast variation of the conductance spectra that become fully gapped as soon as $Z_2$ approaches about 1, suggesting that the wide variety of the observed spectra are favored by the high transparency of the F/S interface in the bilayer. In the latter case, the formation of surface (Andreev) bound states is expected[27].

From Figure 4c we notice a strong dependence of the conductance features on the parameter $r$. The range $48 < r < 58$ corresponds to a thickness $d$ of the ferromagnetic layer of less than 1nm. According to this, the precise fitting of the conductance spectrum in such N-F/S configuration is suitable for direct local measurement of the F-layer thickness. As an example, in Figure 5a we show the different spectra obtained for $r = 51.25$ (i.e. $d \approx 4.3$nm), $r = 52.25$ (i.e. $d \approx 4.4$nm) and $r = 53.25$ (i.e. $d \approx 4.5$nm). To complete the analysis of the conductance spectra dependence on the various parameters, we show in Figure 4d the result obtained by varying $h$ in the range $0.135 < h < 0.145$. This parameter gives direct information about the spin polarization of the ferromagnetic layer. Also in this case, the fast modification of the spectra for small variations of the parameter allows precise estimation of P (see Figure 5b).

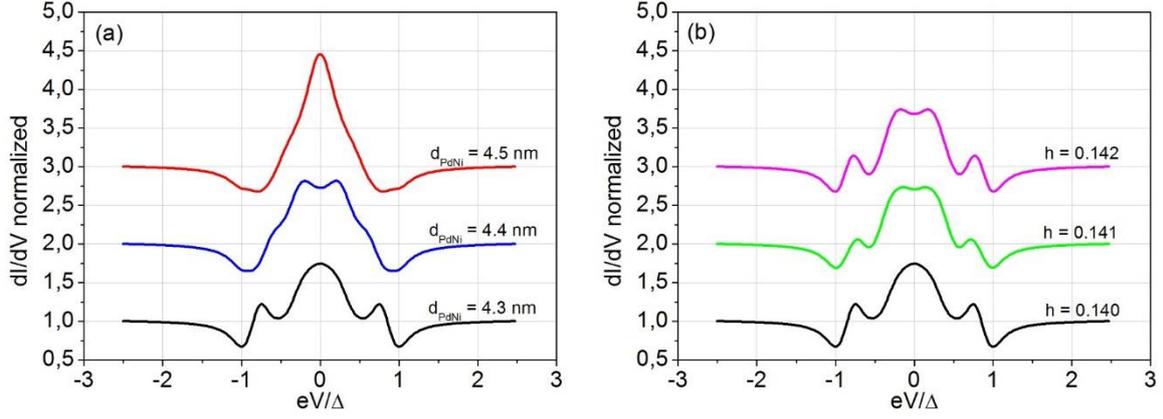

**Figure 5.** Comparison of conductance spectra calculated within our theoretical model for small variations of the parameter (a) $r$ and (b) $h$. The lower (black) spectrum in both panels is the conductance spectrum of Figure 2a calculated with the parameters $Z_1=3.85$, $Z_2=0.35$, $r=51.25$, $h=0.1403$.

We notice that the values of the polarization obtained from the various contacts (and corresponding fittings) of Figure 2 span on a larger range 14.0%< $P$ <14.6%. Due to the high sensibility of the conductance curve on the $h$ value, we can conclude that we are probing small spatial variation of the polarization probed locally by the PCAR setup with the different contacts that have been realized, since we demonstrate that the technique has a resolution better than 0.001 variations on the $h$ values.

*Effective temperature.* As mentioned in the previous section, the numerical simulations need four fitting parameters, namely $Z_1$, $Z_2$, $r$, $h$, while we fixed $\Delta_{Nb} = 1.5$ meV and $T_{eff} = 0.7$ K. According to our experimental setup, in which the point contact inset is directly placed in the liquid helium bath, the temperature of the device is naturally expected to be stabilized at $T = 4.2$ K. On the other hand, the reported conductance features are clearly very sharp in contrast with the thermal smearing expected at T = 4.2 K. Such phenomenon has been already observed for Cu/Nb contacts[7] and it has been ascribed to possible non-equilibrium transport processes in presence of proximity effect at the interface.

We also have to take into account that our experiment (pushing a normal tip on F/S bilayer) realizes a N-F-S device. It has been reported[28] that in such case the ferromagnetic interlayer reduces the Andreev Joule heating favoring an efficient cooling, the cooling power depending on the spin polarization. Moreover, maximum cooling power is expected[29] at $T \approx 0.5\, T_c$ and for bias voltages of about eV/Δ≈1. This process can reasonably affect the local temperature of the N-F-S device under investigation in this work, explaining a discrepancy between $T_{bath}$ and $T_{eff}$. However, the cooling power generally acting on this systems seems to be not enough to cause a temperature reduction of few Kelvin, suggesting that further physical effects should contribute to obtain such experimental observation.

Finally, we also consider the possibility that such small effective temperature could arise from the predicted "extraordinary" temperature dependence of the resonant Andreev reflection peak expected in N/quantum-dot/S systems[30], the tunneling mediated by discrete energy levels being responsible for an anomalous broadening of the conductance peak with respect the thermal one. Experimentally, such behavior has been reported for N/semiconductor/S systems[31] and in N/High-Tc-Superconductors constrictions[32], where the discrete levels could be due to the existence of surface (Andreev) bound states.

In conclusion, we used Bogoliubov-de Gennes formalism to describe N-F-S systems. Experimentally, such configuration has been realized by means of point contact Andreev reflection spectroscopy by pushing a metallic tip on the ferromagnetic side of PdNi/Nb bilayer. Differential conductance spectra for several contacts have been measured at low temperature, showing several different features, all consistently explained within our theoretical model. Moreover, we demonstrated that this setup configuration is suitable to locally measure the ferromagnetic properties (polarization and thickness) of the F-layer with high precision.

AUTHOR INFORMATION

**Corresponding Author**

*Dr. Filippo Giubileo,

Address: CNR-SPIN, via Giovanni Paolo II, 132, 84084 Fisciano (SA), Italy

Tel: +39.089.969329

Fax: +39.089.968817

E-mail: filippo.giubileo@spin.cnr.it

**Author Contributions**

F.G. designed the experiment, performed PCAR measurements and numerical fittings, did data analysis and wrote the paper; F.R. developed theoretical model, performed numerical fittings and contributed to data analysis and paper writing; P.R. contributed to the experiment design, measurements and data analysis; C.C. and C.A. provided the bilayers and their electrical characterization; R.C. contributed to develop the theoretical model; A. DB. and A.P. helped with the data analysis. The manuscript was revised by all authors. All authors have given approval to the final version of the manuscript. ‡These authors contributed equally.

REFERENCES

1. Johnson, P. D. *Rep. Prog. Phys.* **1997,** *60*, 1217-1304.

2. Meservey, R.; Tedrow, P. M. *Phys. Rep.* **1994,** *283*, 173-243.

3. de Jong M. J. M.; Beenakker C. W. J. *Phys. Rev. Lett.* **1995,** *74*, 1657-1660.

4. Andreev A. F. *Sov. Phys. JETP* **1964,** *19,* 1228-1231.

5. Soulen, R. J., Jr.; Byers, J. M.; Osofosky, M. S.; Nadgorny, B.; Ambrose, T.; Cheng, S. F.; Broussard, P. R.; Tanaka, C. T.; Nowak, J.; Moodera, J. S.; Barry, A.; Coey, J. M. D. *Science* **1998,** *282*, 85-88.

6. Upadhyay, S. K.; Palanisami, A.; Louie, R. N.; Buhrman, R. A. *Phys. Rev. Lett.* **1998,** *81*, 3247-3250.

7. Strijkers, G.; Ji, Y.; Yang, F.; Chien, C.; Byers, J. *Phys. Rev. B* **2001,** *63*, 104510.

8. Kant, C. H.; Kurnosikov, O.; Filip, A. T.; LeClair, P.; Swagten, H. J. M.; de Jonge, W. J. M. *Phys. Rev. B* **2002,** *66*, 212403.

9. Nadgorny B.; Soulen, J. R. J.; Osofsky M. S.; Mazin I. I.; G. Laprade, G.; van de Veerdonk R. J. M.; Smite A. A.; Cheng S. F.; Skelton E. F.; Qadri S. B. *Phys. Rev. B* **2000,** *61*, R3788-R3791.

10. Nadgorny, B.; Mazin, I. I.; Osofsky, M.; Soulen, R. J.; Broussard, P.; Stroud, R. M.; Singh, D. J.; Harris, V. G.; Arsenov, A.; Mukovskii, Y. *Phys. Rev. B* **2001,** *63*, 184433.

11. Ji Y.; Chien C.L.; Tomioka Y.; Tokura Y. *Phys. Rev. B* **2002,** *66*, 012410.

12. Raychaudhuri, P.; Mackenzie, A. P.; Reiner, J. W.; Beasley, M. R. *Phys. Rev. B* **2003,** *67*, 020411.

13. Nadgorny B.; Osofsky M.S.; Singh D.J.; Woods G.T.; Soulen R.J.; Lee M.K.; Bu S.D.; Eom C.B. *Appl. Phys. Lett.* **2003,** *82*, 427-429.

14. DeSisto W.J.; Broussard P.R.; Ambrose T.F.; Nadgorny B.E.; Osofsky M.S. *Appl. Phys. Lett.* **2000,** *76*, 3789-3791.

15. Löfwander T.; Grein R.; Eschrig M. *Phys. Rev. Lett.* **2010,** *105*, 207001.

16. Blonder, G.; Tinkham, M.; Klapwijk T. *Phys. Rev. B* **1982,** *25*, 4515-4532.

17. Perez-Willard F.; Cuevas J.C.; Suergers C.; Pfundstein P.; Kopu J.; Eschrig M.; Loehneysen H. v. *Phys Rev B* **2004,** *69*, 140502(R).

18. De Gennes P.G. *Superconductivity of metals and alloys*, Addison-Wesley Publishing Company, 1966.

19. Wexler A. *Proc. Phys. Soc.* **1966,** *89*, 927-941.

20. Sharvin, Y. V. *Zh. Eksp. Teor. Fiz.* **1965,** *48*, 984-985.

21. Maxwell J. C. *Treatise on Electricity and Magnetism*, Clarendon, Oxford, 1904.

22. Goodman B. B.; Kuhn G. *J. Phys. Paris* **1968,** *29*, 240-252.

23. Garwin E. L.; Rabinowitz M. *Appl. Phys. Lett.* **1972,** *20*, 154-156.

24. Mazin I.I.; Golubov A.A. ; Nadgorny B. *J. Appl. Phys.* **2001,** *89*, 7576-7578.

25. Woods G.T.; Soulen J. R.J.; Mazin I.; Nadgorny B.; Osofsky M.S.; Sanders J.; Srikanth H.; Egelhoff W.F.; Datla R. *Phys. Rev. B* **2004,** *70*, 054416.

26. Ashcroft N.W.; Mermin N.D. *Solid State Physics,* Brooks Cole, New York, 1976.

27. Hu C. R. *Phys. Rev. Lett* **1994,** *72*, 1526-1529.

28. Giazotto F.; Taddei F.; Fazio R.; Beltram F. *Appl. Phys. Lett.* **2002,** *80*, 3784-3786.

29. Kawabata S.; Vasenko A.S.; Ozaeta A.; Bergeret S.F.; Hekking F. W. J. arXiv:1407.1977v1

30. Zhu Y.; Sun Q-f.; Lin T-h. *Phys. Rev. B* **2001,** *64*, 134521.

31. Magnée P. H. C.; van der Post N.; Kooistra P. H. M.; van Wees B. J.; Klapwijk T. M. *Phys. Rev. B* **1994,** *50*, 4594-4599.

32. Dagan Y.; Kohen A.; Deutscher G.; Revcolevschi A. *Phys. Rev. B* **2000,** *61*, 7012-7016.